\documentclass[11pt]{article}
\usepackage{epsfig}
\newcommand{\kms}{km~s$^{-1}$}
\newcommand{\farcs}{\hbox{$.\!\!^{\prime\prime}$}}
\newcommand{\hi}{H$\;${\small I}\relax}
\newcommand{\hii}{H$\;${\small II}\relax}
\newcommand{\arcmin}{\hbox{$^\prime$}}
\newcommand{\sun}{\hbox{$\odot$}}
\newcommand{\msun}{$M_{\mbox{\scriptsize \sun}}$}
\newcommand{\gtrsim}{\mathrel{\hbox{\rlap{\hbox{\lower4pt\hbox{$\sim$}}}\hbox{$>$}}}}
\title{{\bf The Large-Scale Structure and Environment of HoII}}
\author{M.~Bureau$^1$, C.~Carignan$^2$\\
\vspace{0.1cm}\\
\normalsize $^1$Sterrewacht Leiden, Postbus~9513, 2300~RA Leiden, Netherlands\\
\normalsize $^2$D\'{e}partement de Physique and Observatoire du Mont
M\'{e}gantic, Universit\'{e} de Montr\'{e}al,\\
C.P.~6128, Succ.~''centre-ville'', Montr\'{e}al, Qu\'{e}bec, Canada H3C~3J7\\
}
\date{}
\begin{document}
\maketitle
%
% Letterhead
%
\def\bull{\vrule height .9ex width .8ex depth -.1ex}
\makeatletter
\def\ps@plain{\let\@mkboth\gobbletwo
\def\@oddhead{}\def\@oddfoot{\hfil\tiny
``Dwarf Galaxies and their Environment'';
International Conference in Bad Honnef, Germany, 23-27 January 2001}%
\def\@evenhead{}\let\@evenfoot\@oddfoot}
\makeatother
%
% Abstract
%
\begin{abstract}\noindent
  Neutral hydrogen VLA D-array observations of the dwarf irregular galaxy
  HoII, a prototype galaxy for studies of shell formation and self-propagating
  star formation, are presented. The large-scale \hi\ morphology is
  reminiscent of ram pressure and is unlikely caused by interactions. A case
  is made for intragroup gas in poor and compact groups similar to the M81
  group, to which HoII belongs. Numerous shortcomings of the supernova
  explosions and stellar winds scenario to create the shells in HoII are
  highlighted, and it is suggested that ram pressure may be able to reconcile
  the numerous observations available.
\end{abstract}
%
% Introduction
%
\section{Introduction}
HoII is a gas-rich dwarf irregular galaxy on the outskirts of the M81 group, at
a distance of 3.2~Mpc ($M_{B_T}=-17.0$~mag). It was one of the first galaxies
outside of the Local Group where the effects of sequential star formation on
the interstellar medium (ISM) were systematically investigated. Puche et al.\ 
(1992) presented high-resolution multi-configuration VLA \hi\ observations,
revealing a complex pattern of interconnected shells and holes. They argued
for a picture where star formation is self-propagating, photoionization,
stellar winds, supernova explosions (SNe), and secondary star formation
shaping the ISM. While we do not wish here to challenge the general relevance
of such scenarios, we will introduce in the next paragraphs numerous problems
they face in the particular case of HoII. Some have been pointed out before,
but others are highlighted here for the first time or are revealed by our
reanalysis of Puche et al.\ (1992) data.
%
% Neutral Hydrogen Distribution of HoII
%
\section{Neutral Hydrogen Distribution of HoII}
We have reanalyzed Puche et al.\ (1992) multi-configuration VLA observations
of HoII, keeping only the D-array data. We produced a continuum-subtracted
naturally-weighted cube cleaned to a depth of 1$\sigma$
(2.75~mJy~beam$^{-1}$), which was used to produce moment maps. The total \hi\ 
map is shown in Figure~1, superposed on an optical image. Undetected before, a
large but faint component extends over the entire northwest half of the
galaxy, encompassing the \hi\ cloud detected by Puche et al.\ (1992). The \hi\ 
on the southeast side of the galaxy is compressed, giving rise to a striking
NW-SE asymmetry, and suggesting that HoII may be affected by ram pressure from
an intragroup medium (IGM). The \hi\ now reaches over 16\arcmin\ (4$R_{25}$),
twice the radius reached previously. The velocity field shows a clear
differentially rotating disk pattern (with a warp) in the inner 7--8\arcmin,
but the kinematics at larger radii is rather disturbed. The largest shells are
clearly visible as peaks in the velocity dispersion. The total \hi\ flux
$F_{\mbox{\scriptsize HI}}=267$~Jy~\kms, corresponding to
$6.44\times10^8$~\msun.
%
% Total HI Map figure (Figure 1)
%
\begin{figure}[t]
\begin{center}
%\vspace*{8cm}
\epsfig{file=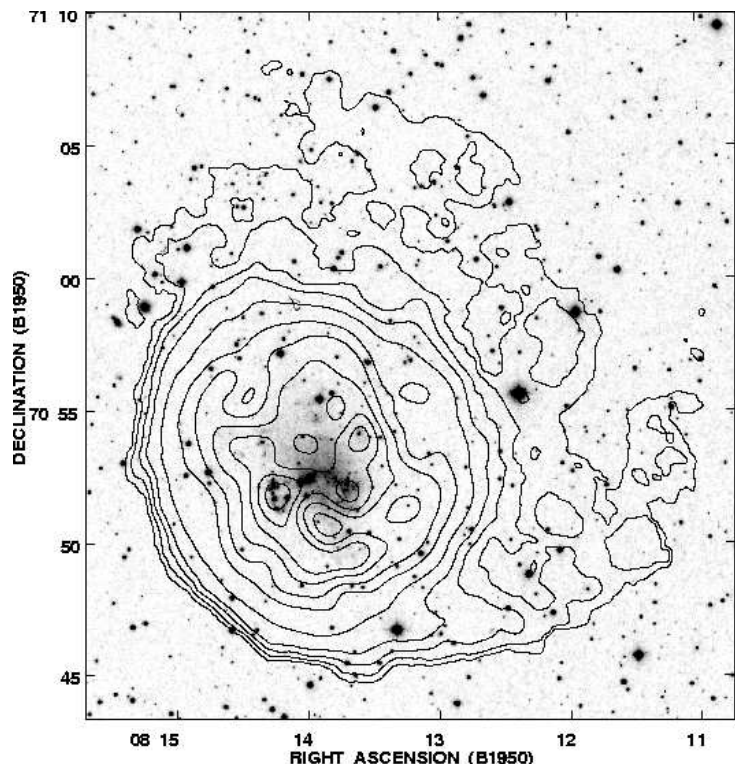,height=10.0cm,clip=}
\caption[]{Total \hi\ map of HoII from the VLA D-array data, superposed on a
  DSS image. Contours are 0.005, 0.015, 0.03, 0.05, 0.10, 0.20, 0.30, 0.45,
  0.60, 0.75, and 0.90 times the peak flux of 8.5~Jy~beam$^{-1}$~\kms\ 
  ($2.10\times10^{21}$~atoms~cm$^{-2}$ or 16.8~\msun~pc$^{-2}$). The beam is
  $66\farcs7\times66\farcs7$.}
\end{center}
\end{figure}
%
% Environment of HoII
%
\section{Environment of HoII}
The morphology of the \hi\ in Figure~1 is reminiscent of ram pressure, but it
could also be caused by interactions. HoII lies 475~kpc in projection from the
center of the M81 group, defined by M81, M82, and NGC~3077. At the group
velocity dispersion of 110~\kms\ (e.g.\ Huchra \& Geller 1982), it would take
HoII a third of a Hubble time to reach the center of the group. However, HoII
appears to be part of subsystem of three dwarf irregular galaxies to the
northwest of the group's core, along with Kar~52 (M81~Dwarf~A) and UGC~4483.
If HoII is interacting, it must be with one of these two galaxies. Both Kar~52
and UGC~4483 are much smaller and fainter than HoII and have an irregular
optical morphology, but none shows obvious signs of interactions (Bremnes,
Binggeli, \& Prugniel 1998). In \hi, Kar~52 displays a lumpy ring with little
rotation, incomplete to the NW (Sargent, Sancisi, \& Lo 1983), and UGC~4483
shows a peaked distribution with a faint envelope extended NW-SE (WHISP
database). The distance to HoII is identical to that of UGC~4483, but also to
that of NGC~2403 and DDO~44 to the SW. This suggests that HoII belongs to the
NGC~2403 subgroup, along with Kar~52, UGC~4483, NGC~2365, and DDO~44.
Karachentsev et al.\ (2000) show that the NGC~2403 subgroup is nearer to us by
0.5~Mpc but has a larger radial velocity, suggesting that it is moving towards
the M81 group at 110-160~\kms.  Observations of the environment of HoII
therefore do not support interactions as a likely mechanism to shape its
large-scale structure, and considerations of the M81 and NGC~2403 subgroup as
a whole suggest that HoII could have a large velocity relative to a putative
IGM. Ram pressure must therefore be considered a serious candidate to explain
the \hi\ morphology of HoII. \hi\ observations of the entire region around
HoII, Kar~52, and UGC~4483 should help clarify this issue.
%
% IGM and X-Rays in Small Groups
%
\section{IGM and X-Rays in Small Groups}
The condition for ram pressure stripping can be written as the balance between
the pressure exerted by an ambient medium and the restoring force of a
galactic disk. The ISM will be stripped if
\begin{equation}
\rho_{\mbox{\tiny IGM}}v^2>2\pi G\Sigma_{tot}\Sigma_{g}
\end{equation}
(Gunn \& Gott 1972), where $\rho_{\mbox{\tiny IGM}}$ is the IGM density, $v$
the relative velocity of the galaxy with respect to the IGM, and
$\Sigma_{tot}$ and $\Sigma_{g}$ the total and ISM surface densities,
respectively. Taking $v\approx\sigma\approx110$~\kms\ and $\Sigma_{tot}$ and
$\Sigma_{g}$ (corrected for other gaseous species) at the first significantly
disturbed contour in Figure~1, we derive a critical density for ram pressure
$\rho_{\mbox{\tiny IGM}}\gtrsim2.3\times10^{-5}$~atoms~cm$^{-3}$.

A virial mass of $1.13\times10^{12}$~\msun\ is derived from the five most
prominent members of the M81 group (Huchra \& Geller 1982). Spreading 1\% of
this mass in a sphere just enclosing HoII, we obtain a mean density of
$1.0\times10^{-6}$~atoms~cm$^{-3}$, 25 times too little for stripping.
Although it suffers from many shortcomings, this number provides a benchmark
with which to compare more sophisticated calculations. The IGM is likely to be
more concentrated and clumpy, and the encounter may not be exactly
``face-on'', but since the group velocity dispersion estimate is based only on
a few large galaxies, it is probably an underestimate, and it is in any case
unlikely that the group is virialized at the distance of HoII, making its
three-dimensional velocity highly unconstrained. If HoII is bound to the M81
group, then the virial mass adopted is severely underestimated. These factors
can easily bring the required and derived IGM densities within a factor 2--3
of each other. More convincingly, typical parameters for poor groups are
$R_{vir}\sim0.5h^{-1}$~Mpc and $M_{vir}\sim0.5-1\times10^{14}h^{-1}$~\msun\ 
(Zabludoff \& Mulchaey 1998), of which only 10--20\% is associated with
individual galaxies, leading to a mean density for the remaining matter of
$\sim3\times10^{-3}$~atoms~cm$^{-3}$ (within the virial radius). If only 1\%
of this is ordinary interacting baryonic matter, then its density is
sufficient to strip galaxies like HoII of their ISM. This is promising, since
on scales of the virial radius, the dominant baryonic mass component in groups
is the IGM (Mulchaey 2000). In fact, Zabludoff \& Mulchaey (1998) report X-ray
gas masses of $1\times10^{12}h^{-5/2}$~\msun\ for their groups, leading to
mean densities for the hot gas of $\sim7\times10^{-5}$~atoms~cm$^{-3}$ within
the virial radius.

The total X-ray luminosity in groups does not correlate with either the total
number of galaxies or the optical luminosity, but it does correlate with the
group velocity dispersion and gas temperature. A common fit to cluster and
compact group data yields, for $\sigma=110\pm10$~\kms,
$L_X=10^{39.6\pm1.7}$~erg~s$^{-1}$ and $T_{\mbox{\tiny
    IGM}}=10^{-0.91\pm0.13}$~keV (Ponman et al.\ 1996). A similar correlation
for loose groups alone yields $L_X=10^{40.5\pm3.6}$~erg~s$^{-1}$ and
$T_{\mbox{\tiny IGM}}=10^{-0.48\pm0.10}$~keV (Helsdon \& Ponman 2000). The
large errors on $L_X$ are probably related to the wind injection histories of
the groups, which also lead to rather flat surface brightness profiles. There
are also indications that the correlations for groups and clusters may be
different, so both $L_X$ and $T_{\mbox{\tiny IGM}}$ are probably
underestimates. There can thus be a substantial amount of hot gas in groups
like the M81 group.

Another issue of interest is the survival of any stripped gas in the hot IGM.
Following Cowie \& McKee (1977), the evaporation timescale for a typical cloud
($n\approx1$~cm$^{-3}$, $R\approx10$~pc) embedded in an IGM at the
temperatures mentioned above is $6.2\times10^5$ to $2.9\times10^7$~yr. The
disturbed ISM in HoII extends over $7-8\arcmin$ in the radial direction. At a
velocity of 110~\kms, it takes HoII about $6\times10^7$~yr to cross that
distance. Given the strong dependence of the evaporation on the assumed
properties of the clouds and IGM, the timescales calculated seem consistent
with the observations. It is also possible to show that, in the conditions of
interest here, cooling and viscous stripping (Nulsen 1982) are negligible
compared to evaporation. At the IGM density required for ram pressure
stripping, the cooling timescale is $6.0\times10^9-5.9\times10^{10}$~yrs and
the timescale for (complete) viscous stripping of our typical cloud is
$5.2\times10^9$~yr.
%
% Creation of Shells and Supershells
%
\section{Creation of Shells and Supershells}
Over 50 \hi\ holes were cataloged in HoII by Puche et al.\ (1992), who
obtained estimates of their kinematics ages, previously enclosed \hi\ masses,
and creation energies. Many correlations between these quantities are
presented, arguing for a formation of the holes through SNe and stellar winds,
but many are truly a reflection of the properties of HoII rather than of the
formation mechanism of the holes. Also, H$\alpha$ emission does not
preferentially fill small holes or trace the edges of large ones. A similar
picture emerges from far-ultraviolet observations (Stewart et al.\ 2000).
Furthermore, the shells are devoid of hot gas, and X-ray emission is not
preferentially associated with \hii\ regions or \hi\ holes; only a mixed bag
of objects is observed (Kerp \& Walter 2001). The SN rate derived from radio
continuum observations agrees well with that derived from the \hi\ shells
(Tongue \& Westpfahl 1995), but the energy is deposited in the central regions
of HoII only, hardly helping to explain how the entire web of interconnected
shells formed. Furthermore, in most cases where useful limits are derived,
stellar clusters expected to remain from massive star formation episodes are
simply not seen (Rhode et al.\ 1999). Multi-wavelength observations thus pose
a challenge to SNe and stellar winds scenarios for the formation of the shells
in HoII, particularly in the outer parts of the disk where no star formation
appears to be taking place.

Other mechanisms exist to explain the formation of the shells and reconcile
models with observations (see Rhode et al.\ 1999).  In particular, SN may not
be spherically expanding in a uniform ISM, as assumed, and the initial mass
function could be very top-heavy. Gamma-ray bursts may also be more efficient
at creating holes and shells. All these mechanisms, however, still require
massive star formation in the outer parts of the disk. A fractal \hi,
overpressured \hii\ regions, external ionization sources, and/or high-velocity
clouds can bypass this requirement.

Here, we would like to suggest that ram pressure can provide yet another
solution to the problem of shell formation in HoII. Ram pressure can create
holes in an \hi\ disk where local minima in the surface density exist. Thus,
it can provide an efficient mechanism to enlarge pre-existing holes, created
by SNe or otherwise, and it can explain the overestimated energy requirements
(or lack of observational signature) from SNe and stellar wind scenarios.
Obviously, however, shell formation through ram pressure should be properly
modeled before making further claims. Ram pressure should be easy to
distinguish from internal, pressure-driven events such as SNe and stellar
winds, as the shells will have a ``bullet-hole'' geometry similar to that
caused by high-velocity cloud collisions (e.g.\ Tenorio-Tagle 1980). Of
course, in the case of HoII, a direct proof of a sufficiently dense IGM must
also be found before any ram pressure model can be taken seriously.
%
% References
%
\begin{description}{} \itemsep=0pt \parsep=0pt \parskip=0pt \labelsep=0pt
\item {\bf References}

{\small
\item Bremnes, T., Binggeli, B., \& Prugniel, P.\ 1998, A\&AS 129, 313
\item Cowie, L.\ L., McKee, C.\ F.\ 1977, ApJ 211, 135
\item Gunn, J.\ E., \& Gott, J.\ R.\ III 1972, ApJ 176, 1
\item Helsdon, S.\ F., \& Ponman, T.\ J.\ 2000, MNRAS 315, 356
\item Huchra, J.\ P., \& Geller, M.\ J.\ 1982, ApJ 257, 423
\item Karachentsev, I.\ D., et al.\ 2000, A\&A 363, 117
\item Kerp, J., \& Walter, F.\ 2001, in {\em X-ray Astronomy 2000}, eds.\ R.\ Giacconi, L.\ Stella, \& S.\ Serio, in press.
\item Mulchaey, J.\ S.\ 2000, ARA\&A 38, 289
\item Nulsen, P.\ E.\ J.\ 1982, MNRAS 198, 1007
\item Ponman, T.\ J., Bourner, P.\ D.\ J., Ebeling, H., \& B\"{o}hringer, H.\ 1996, MNRAS 283, 690
\item Puche, D., Westpfahl, D., Brinks, E., \& Roy, J.-R.\ 1992, AJ 103, 1841
\item Rhode, K.\ L., Salzer, J.\ J., Westpfahl, D.\ J., \& Radice, L.\ A.\ 1999, AJ 118, 323
\item Sargent, W.\ L.\ W., Sancisi, R., Lo, K.\ Y.\ 1983, ApJ 265, 711
\item Stewart, S.\ G., et al.\ 2000, ApJ 529, 201
\item Tenorio-Tagle, G.\ 1980, A\&A 88, 61
\item Tongue, T.\ D., \& Westpfahl, D.\ J.\ 1995, AJ 109, 2462
\item Zabludoff, A.\ I., \& Mulchaey, J.\ S.\ 1998, ApJ 496, 39
}
\end{description}
\end{document}